\documentstyle[hip-artc]{article}  

\volnumber{5}  \edyear{1997}  \frompage{000} \topage{000}                
\recrevdate{1 January 1996; revised version 1 January 1997} 			 
\title{$^7$Be($p,\gamma$)$^8$B and the high-energy solar neutrino flux} 
\authors{
{\twerm Attila Cs\'ot\'o %
}\\[2.812mm]
{\normalsize
Theoretical Division, Los Alamos National Laboratory, \\ 
Los Alamos NM 87545, USA\\[0.2ex] 
}}
 
\abstract{ The importance of the $^7$Be($p,\gamma$)$^8$B reaction in 
predicting the high-energy solar neutrino flux is discussed. I 
present a microscopic eight-body model and a potential model for 
the calculation of the $^7$Be($p,\gamma$)$^8$B cross section.} 
 
\begin{document}
 
\maketitle
 
\section{Introduction}
 
Today the neutron halo structure in some nuclei near the neutron drip 
line is well established. The proton drip line is, however, less 
well understood. The first (and so far still best) candidate for a 
proton halo structure was $^8$B. In this paper I show how important 
it is to understand the structure of $^8$B in order to be able to 
give precise predictions for the high-energy solar neutrino flux.

The high level of sophistication achieved in standard electroweak and 
stellar physics enables us to understand in detail the synthesis and 
evolution of the elements of the Universe in the Big Bang and in stars. 
The astronomical observations and theoretical predictions are in a very good 
general agreement with each other, with a few notable exceptions. Perhaps 
the most notorious of these exceptions is the solar neutrino problem. 
Despite thirty years of extensive experimental and theoretical work, the 
predicted solar neutrino fluxes are still in sharp disagreement 
with measurements. The solar neutrino measurements strongly suggest that 
the problem cannot be solved within the standard electroweak and 
astrophysical theories. Thus, the solar neutrino problem constitutes the 
strongest (and so far almost only) evidence for physics beyond the Standard 
Model \cite{Bahcall}.

Whatever the solution of the solar neutrino problem turns out to be, it 
is of paramount importance that the input parameters of the underlying 
electroweak and solar theories rest upon solid ground. 
The most uncertain nuclear input parameter in standard solar
models is the low-energy $^7$Be($p,\gamma$)$^8$B radiative
capture cross section. This reaction produces $^8$B in the
Sun, whose $\beta^+$ decay is the main source of the 
high-energy solar neutrinos. Many present and future solar neutrino 
detectors are sensitive mainly or exclusively to the $^8$B neutrinos 
(Fig.\ 1). 

\begin{figure}[bt]
\begin{center}
\leavevmode
\epsfysize=5cm
\epsfbox{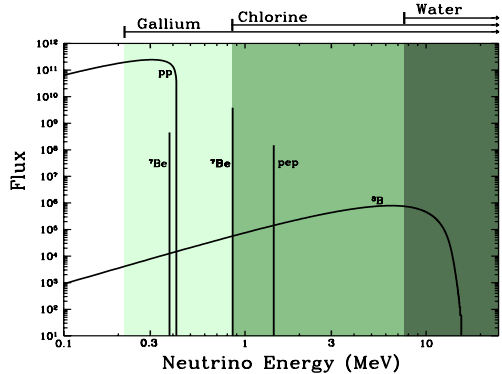}
\end{center}
\vspace*{-0.7cm}
\caption[]{
Solar neutrino fluxes predicted by the solar model of Bahcall 
\cite{Bahcall}. The thresholds of the various detectors are indicated.
}
\label{fig1}
\vspace*{-0.5cm}
\end{figure}

The predicted $^8$B neutrino flux is proportional to 
$S_{17}$, the $^7$Be($p,\gamma$)$^8$B astrophysical $S$ factor
($S(E)=\sigma(E)E\exp[2\pi\eta(E)]$), at solar energies ($E_{cm}=20$ keV).  
Thus, the value of $S_{17}$(20 keV) is a crucial input parameter 
in solar models.

Currently there is a considerable confusion concerning the
value of $S_{17}(0)$. The six direct capture measurements
performed to date give $S_{17}(0)$ between 15 eVb and 40 eVb,
with a weighted average of $22.2\pm 2.3$ eVb, while a recent 
Coulomb dissociation measurement give $S_{17}(0)=16.7\pm 3.2$ eVb.
The theoretical predictions for $S_{17}(0)$ also have a huge
uncertainty, as the various models give values between 16 eVb
and 30 eVb. For an experimental and theoretical summary, see \cite{Langanke}. 

In order to set tighter theoretical limits on the value of $S_{17}$, 
the nonresonant part of the $^7$Be($p,\gamma$)$^8$B reaction has been 
studied in a microscopic eight-body model \cite{micr} and in a $^7$Be$+p$ 
potential model \cite{pot}, respectively.

\section{$^7$Be($p,\gamma$)$^8$B cross section constrained by A=7 and 8 
observables}
The relatively low temperature (on nuclear scale) of our Sun means that all 
charged-particle reactions in the energy-generating solar p-p chain 
take place well below the Coulomb barrier. In such cases the radiative 
capture cross section gets contributions 
almost exclusively from the external nuclear regions ($r>6-8$ fm). At such 
distances the scattering ($^7$Be+$p$) and bound state ($^8$B) wave functions 
are fully determined, provided the scattering phase shifts and bound state 
asymptotic normalizations are known. At solar energies the 
phase shifts coincide with the (almost zero) hard sphere phase shifts, while 
the bound state wave function in the external region behaves like $\bar c
W^+(kr)/r$, where $W^+$ is the Whittaker function and $\bar c$ is the 
asymptotic normalization. So the only unknown parameters 
that govern the solar radiative capture reactions, like 
$^7$Be($p,\gamma$)$^8$B, are the $\bar c$ values. The $\bar c$ normalization 
depends mainly on the effective $^7$Be--$p$ interaction radius. 
A larger radius results in a lower Coulomb barrier, which leads to a higher 
tunneling probability into the external region, and thus to a higher cross 
section. Our aim is to constrain the theoretical interaction radius by 
some properties of A=7 and 8 observables. 

In the first model \cite{micr} an eight-body three-cluster wave function 
is used, which is  variationally converged and virtually complete in 
the $^4$He+$^3$He+$p$ cluster model space. We find that the low-energy 
astrophysical $S$ factor is linearly correlated with the quadrupole 
moment of $^7$Be (Fig.\ 2a). A range of parameters is found where the 
most important $^8$B, $^7$Be and $^7$Li properties are reproduced 
simultaneously; the corresponding $S$ factor at zero energy is $25-26.5$ 
eVb. 

\begin{figure}[bt]
\begin{center}
\leavevmode
\epsfysize=4.0cm
\epsfbox{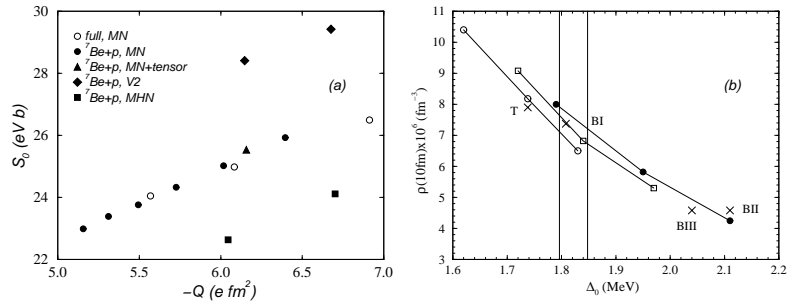}
\end{center}
\vspace*{-0.7cm}
\caption[]{
Correlation between (a) the zero-energy astrophysical $S$ factor 
of the $^7$Be($p,\gamma$)$^8$B cross section and the quadrupole moment 
of $^7$Be in the microscopic eight-body model \cite{micr}; and (b) the 
valence-proton density at 10 fm and the Coulomb displacement energy in 
the potential model \cite{pot}. 
In Fig.\ (a) the results of several calculations, using various $N-N$ 
interactions and model spaces, are shown. In Fig.\ (b) the three lines 
join points obtained for different values of the diffuseness of the 
potential. The crosses show the results of potentials that are previously 
used in the literature. The region between the vertical lines are the 
range of values allowed by the displacement energy. 
}
\label{fig2}
\vspace*{-0.5cm}
\end{figure}

Our second model \cite{pot} describes the $^7$Be($p,\gamma$)$^8$B process by 
assuming a local potential between a structureless $^7$Be and $p$. The 
relationship between the Coulomb displacement energy for the A=8, J=2$^{ + }$, 
T=1 state and the low-energy astrophysical $S$ factor for the $^{7}$Be($p,
\gamma$)$^{8}$B reaction is studied. The displacement energy is interpreted 
in a particle-hole model. The dependence of the particle displacement energy 
on the potential well geometry is investigated and is used to relate the 
particle displacement energy to the rms radius and the asymptotic 
normalization of the valence proton wave function in $^{8}$B. We find 
that the theoretical Coulomb displacement energy strongly 
constrains the valence-proton density, and thus the asymptotic normalization, 
in the external region (Fig.\ 2a). The asymptotic normalization is used to 
calculate $S_{17}$. The model predicts $S_{17}$ to be $24.5\pm2.9$ eVb at 
zero energy. 

\begin{figure}[tb]
\begin{center}
\leavevmode
\epsfysize=5.8cm
\epsfbox{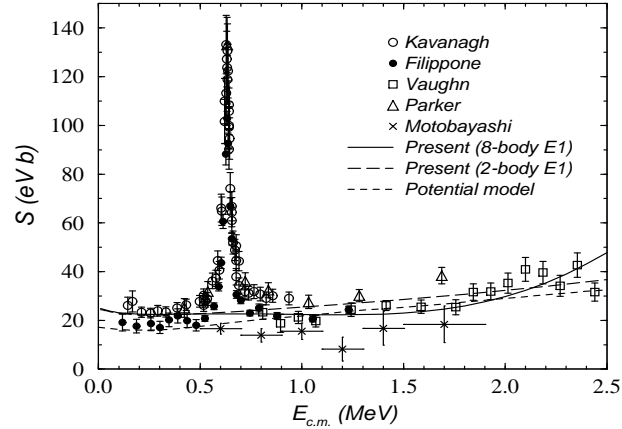}
\end{center}
\vspace*{-0.7cm}
\caption[]{
Astrophysical $S$ factor for the $^7$Be($p,\gamma$)$^8$B reaction \cite{offsh}. 
The symbols show the various experimental data. The solid and long-dashed lines 
are the $E1$ components of the $S$ factors in the eight-body model of 
\cite{micr} with and without antisymmetrization in the electromagnetic 
transition matrix, respectively. The short-dashed line is the result of 
a typical potential model.
}
\label{fig3}
\vspace*{-0.5cm}
\end{figure}

The energy dependence of the $S$ factor is also studied \cite{offsh} with 
the aim to understand the apparent disagreement between the predictions 
of the microscopic eight-body model \cite{micr} and potential models. It 
is found that off-shell effects, like antisymmetrization and core-excitation 
and deformation, significantly influence the energy dependence of the $S$ 
factor. The proper treatment of these effects results in a virtually flat 
E1 component of the $S$ factor at $E_{cm}=0.3-1.5$ MeV in the present 
eight-body model (Fig.\ 3). Off-shell effects can cause 15--20\% changes 
in the value of $S_{17}(0)$ extrapolated from high-energy ($E_{cm}>0.7$ 
MeV) data. 

\section{Conclusion}
The zero-energy $S$ factors, predicted by our models, are 
slightly higher than but compatible with $S_{17}(0)=22.4\pm 2.1$ eVb, 
used in standard solar models \cite{Bahcall}.

\vspace*{0.2cm}
\noindent This work was performed under the auspices of the U.S.\ Department 
of Energy, and was also supported by OTKA Grant F019701.

\vfill\eject

\begin{thebibliography}{99}  
\bibitem{Bahcall}J~N. Bahcall, {\it Neutrino Astrophysics} (Cambridge 
University Press, 1989). 
\bibitem{Langanke} K. Langanke, in {\it Solar Modeling} (World 
Scientific, Singapore, 1995). 
\bibitem{micr} A. Cs\'ot\'o, K. Langanke, S.~E. Koonin, and T.~D. 
Shoppa, {\it Phys. Rev.} {\bf C52} (1995) 1130.   
\bibitem{pot} B.~A. Brown, A. Cs\'ot\'o, and R. Sherr, {\it Nucl. Phys.} 
{\bf A597} (1996) 66.
\bibitem{offsh} A. Cs\'ot\'o, {\it Phys. Lett.} {\bf B394} (1997) 247.
\end{thebibliography}
\end{document}